# Raman Spectral Indicators of Catalyst Decoupling for Transfer of CVD Grown 2D Materials

Patrick R. Whelan[a], Bjarke S. Jessen[a,b], Ruizhi Wang[c], Birong Luo[a], Adam C. Stoot[a], David M. A. Mackenzie[a], Philipp Braeuninger-Weimer[c], Alex Jouvray[d], Lutz Prager[e], Luca Camilli[a], Stephan Hofmann[c], Peter Bøggild[a,b], and Timothy J. Booth[a]*

*E-mail: tim.booth@nanotech.dtu.dk

[a]DTU Nanotech, Technical University of Denmark, Ørsteds Plads 345C, DK-2800, Denmark
[b]Center for Nanostructured Graphene (CNG), Technical University of Denmark, DK-2800, Denmark
[c]Department of Engineering, University of Cambridge, Cambridge, CB3 0FA, United Kingdom
[d]AIXTRON Ltd, Anderson Road, Buckingway Business Park, Swavesey, Cambridge, CB24 4FQ, United Kingdom
[e]Leibniz-Institut für Oberflächenmodifizierung e.V. Permoserstrasse 15, 04318 Leipzig, Germany

**Abstract**

Through a combination of monitoring the Raman spectral characteristics of 2D materials grown on copper catalyst layers, and wafer scale automated detection of the fraction of transferred material, we reproducibly achieve transfers with over 97.5% monolayer hexagonal boron nitride and 99.7% monolayer graphene coverage, for up to 300 mm diameter wafers. We find a strong correlation between the transfer coverage obtained for graphene and the emergence of a lower wavenumber $2D^-$ peak component, with the concurrent disappearance of the higher wavenumber $2D^+$ peak component during oxidation of the catalyst surface. The 2D peak characteristics can therefore act as an unambiguous predictor of the success of the transfer. The combined monitoring and transfer process presented here is highly scalable and amenable for roll-to-roll processing.

# 1. Introduction

Transfer of graphene and other 2D materials from catalytic growth substrates is typically performed by complete dissolution of the copper catalyst layer[1–4] – however, this can negatively impact the properties of the transferred materials due to contamination by residues of the catalyst and etching solution.[5,6] Furthermore, the recovery of dissolved catalyst or disposal has consequences for the cost and environmental impact of such processes. Transfer of 2D materials has also been achieved by a variety of techniques that do not require destruction of the catalyst, including mechanical peeling,[7–9] electrochemical delamination by hydrogen evolution,[10–13] interfacial oxidation,[14] and a range of intercalation based techniques.[15–18] The common element between all of these transfer methods is the requirement to decouple graphene from the catalyst layer without the introduction of mechanical damage or contamination.

Here we show that the degree of decoupling of graphene from a copper catalyst layer can be measured, and the subsequent coverage of transferred graphene on a target substrate accurately predicted by monitoring the evolution of the graphene 2D peak characteristics during the decoupling process. We use a water-based catalyst oxidation at elevated temperature and subsequent mechanical peeling as a model transfer system to demonstrate the utility of this technique. Automated large-scale optical microscopic mapping of the transferred materials enables us to measure the precise resulting coverage and extent of mechanical damage and other possible inhomogeneities - including second and third layer growths and polymer residues. This allows us to directly correlate the extent of the graphene transferred with the Raman 2D peak spectral characteristics.

## 2. Experimental Section

### 2.1. 2D materials growth

Graphene was synthesized by chemical vapor depositiong (CVD) in a cold wall system (Aixtron Black Magic Pro 4") on either a 1.5 μm film of sputtered Cu supported by an oxidized Si wafer or on electropolished Cu foil (25 μm thick, polished in 20 % phosphoric acid). 300 mm graphene wafers were synthesized using an Aixtron BM300T cold wall CVD system on commercially available Cu/SiO$_2$/Si wafers. The CVD growth of graphene on Cu, which follows published recipes,[19–21] consists of an initial annealing phase in a H$_2$/Ar atmosphere and a growth phase in which a CH$_4$ precursor is introduced in the chamber. Finally, the sample was cooled down to room temperature with an Ar flow. The graphene domain size is on the order of 10-20 μm analyzed from scanning electron microscopy (SEM) images of incomplete growth.

Growth of hexagonal boron nitride (hBN) was performed on electropolished Cu foil. A commercial tube furnace (MTIx) was used to bring the sample to 900°C before adding a flow of 3 sccm argon-bubbled borazine, 15 sccm hydrogen and 300 sccm argon over the sample for 15 minutes at 60 torr. Only the argon flow was maintained during cool-down to room temperature.

### 2.2 Transfer of 2D materials

The model transfer method, adapted from that presented by Yang et al.[7], is outlined in Fig. 1a-f. Cu was oxidized beneath the graphene layer by immersing samples in deionized water. The graphene/Cu sample was washed in isopropyl alcohol (IPA) and dried under a nitrogen flow. Polyvinyl alcohol (PVA) solution (5 g PVA and 1 g glycerol per 100 mL DI water) was drop coated onto the sample and dried on a hotplate at 80°C. Glycerol was added as plasticizer[22] to the PVA solution to soften the polymer. Thermal release tape (TRT) was used to support mechanical peeling of PVA/graphene from the Cu substrate and the TRT/PVA/graphene stack was subsequently pressed onto the target substrate at 130°C to release the TRT support. The PVA/graphene was left on the target substrate for 5 minutes at 130°C before placement in water at 40°C for 3 h to dissolve the PVA. We also performed additional transfers using a roll-to-roll silica-coated polymer film[23], 250 mm wide, used as a flexible single-layer gas barrier in applications as a substitute for TRT. Finally, the graphene/target substrate was rinsed with IPA and dried under a nitrogen flow. The exact same steps were followed for transfer of hBN from Cu onto $SiO_2$. The target substrate is silicon with 90 nm $SiO_2$ unless stated otherwise.

*2.3 Characterization*

Raman measurements were conducted using a Thermo DXRxi Raman Spectrometer with a 455 nm laser. Raman spectra of Cu/graphene samples during water oxidation were acquired by removing the sample from water for measurements at specific time steps before re-immersion. The Raman spectra presented from Cu/graphene samples are the average of at least 16 Raman spectra acquired from the same region on the sample. Raman peak intensities and positions were found by fitting individual Raman peaks of the average spectrum to Lorentzian functions with error bars representing the standard error on the fit. The hBN peak position on $SiO_2$ was determined by fitting to a single Lorentzian function.

Optical images were acquired with either a Nikon Eclipse LN200 or a Nikon Eclipse LV150N. To determine the coverage of graphene, we start by calculating the wavelength-dependent contrast of graphene.[24] Using the red, green, and blue (RGB) spectral response functions of the CCD sensor, we can obtain the numerical RGB profile for every pixel corresponding to silicon oxide, single layer graphene, and bilayer graphene.[14] Pixels which do not fall into any of these three categories are labelled as 'other' (three or more layers of graphene and other sources of contrast such as polymer residues), and the coverage values presented in this paper thus represent a lower bound, as pixels containing both graphene and residues will only count towards the coverage of residues. The graphene coverage values presented here are the average of coverage data from at least 5 images of 1440 by 1070 pixels at a resolution of 0.241 µm/pixel with error bars representing the standard error on the mean. The coverage of a material is determined by dividing the total number of pixels corresponding to that material (in practice single layer or single and bilayer graphene) by the number of pixels with SiO$_2$, single and bilayer graphene contrast. Similarly, a coverage map of hBN can be made by calculating the relative wavelength-dependent contrast.[25]

Transmission electron microscopy (TEM) was performed with a Tecnai T20 G2 TEM operated at 200 kV. Lamellas for TEM analysis were prepared by focused ion beam (FIB) liftouts onto TEM grids in a Helios NanoLAB 600 SEM equipped with an ion beam gun and an Oxford Omniprobe micro manipulator. A thin layer of Pt was deposited on the area of the extracted Cu/graphene sample as a protection layer from ion beam damage while trimming the lamella down to a thickness below 100 nm.

X-ray photoelectron spectroscopy (XPS) was performed using a commercial Thermo Scientific K-Alpha with a monochromized Al K$_\alpha$ source.

## 3. Results

CVD-grown graphene on copper samples are immersed in water for fixed durations from 0 to 150 minutes (Fig. 1a) before a PVA based mechanical peeling process is applied to transfer the graphene to SiO$_2$ substrates (Fig. 1b-f). Raman spectra acquired during water oxidation show a redshift in the G peak from 1585 to 1579 cm$^{-1}$, along with a characteristic splitting of the 2D peak into 2D$^+$ and 2D$^-$ components at 2706 cm$^{-1}$ and 2685 cm$^{-1}$, respectively (Fig. 1g,h). The 2D$^+$ peak intensity decreases with extended oxidation in water, with the 2D$^-$ peak intensity simultaneously increasing. A similar shift in the 2D peak position after Cu oxidation in ambient air[26] and in water[27] has previously been reported. Six Raman peaks appear in the range from 200 cm$^{-1}$ to 800 cm$^{-1}$ after the sample has been immersed in water (Supplementary Fig. S1) with the intensity increasing after longer oxidation time. All the six peaks can be identified as Cu$_2$O and CuO peaks,[28–30] indicating that oxidation of the Cu surface is taking place under the graphene. Gigapixel optical microscopic mapping of the transferred graphene layers (Fig. 1i-k) shows distinct regions which can be assigned, based on pixel contrast to either $SiO_2$, single layer graphene (SLG), bilayer graphene (BLG) or 'other' corresponding to both three or more layers of graphene and other sources of contrast. '$SiO_2$' areas represent regions where graphene was not transferred, where graphene was damaged/folded during transfer, or missing graphene regions in the original growth. Bilayer and 'other' contrasts are obtained from multilayer regions resulting from the growth (Fig. 1k), and from folded and rolled up monolayer regions caused by mechanical damage that occurs to the graphene during transfer (Fig. 1i, j).

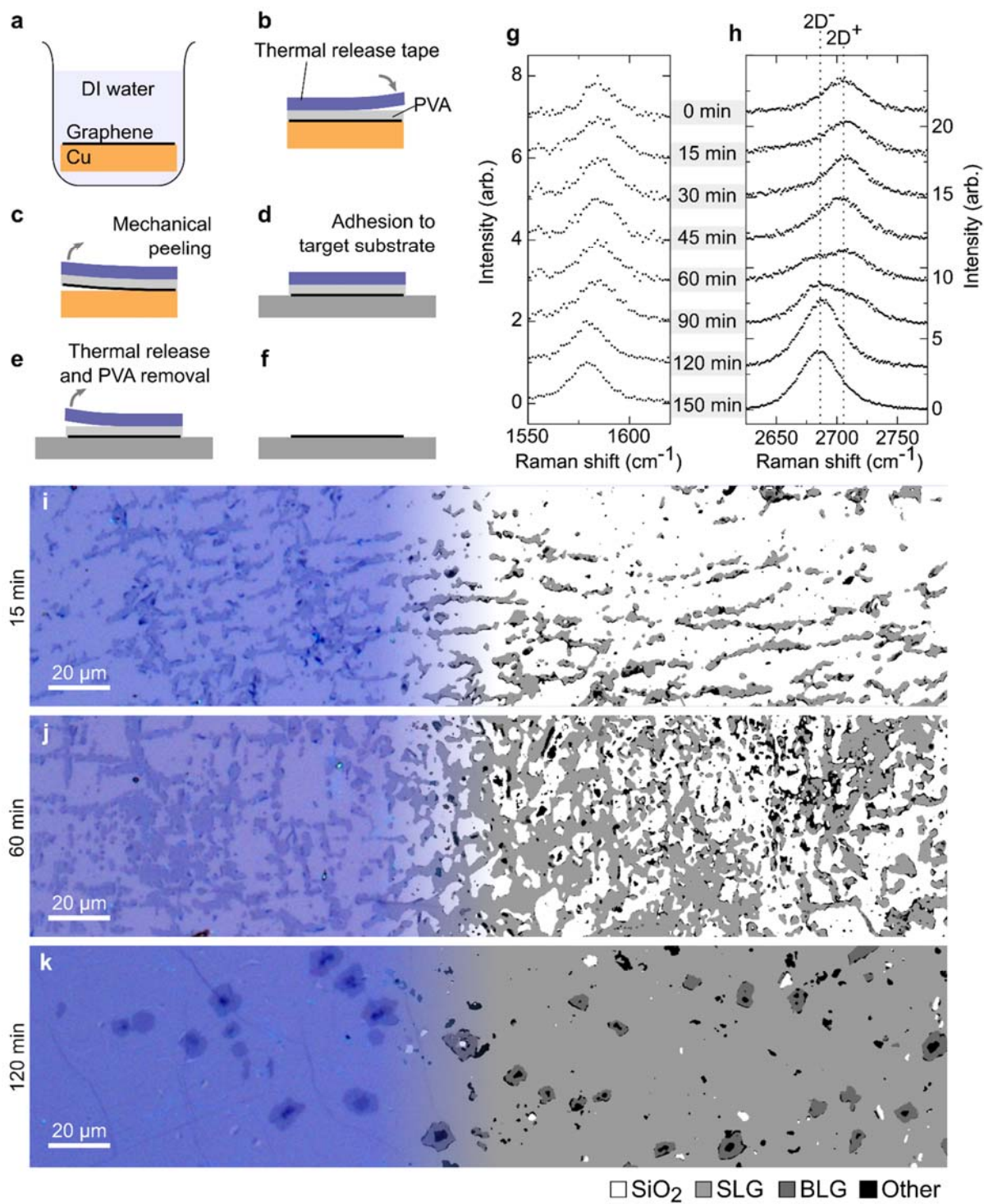

**Fig. 1.** (a) Graphene on a copper catalyst layer is oxidized in deionized water at elevated temperature. (b) PVA is applied by spin coating and a thermal release tape applied as a support. (c) The graphene and PVA are mechanically peeled from the substrate, and (d) adhered to a target substrate. (e) Elevated temperature is used to remove the thermal release tape support layer, and (f) DI water is used to remove the PVA. (g) Raman G and (h) 2D peaks acquired from Cu/graphene samples for different oxidation times in 40°C water. Intensities are normalized to the G peak intensity. A small redshift in G peak with oxidation time is accompanied by the suppression of the $2D^+$ peak at 2706 cm$^{-1}$ and increase in intensity of the $2D^-$ peak at 2685 cm$^{-1}$. (i-k) White light optical microscopy (left) and coverage images (right) of graphene transferred to SiO$_2$ after immersion for 15, 60, and 120 minutes, respectively, in 40°C water.

The coverage of graphene obtained by this technique increases with increasing oxidation time in water from 0% to nearly 100% (Fig. 2a). More importantly, the characteristics of the Raman 2D peak are strongly correlated with the coverage of graphene obtained by transfer. We define the ratio

$$\Theta = \frac{I(2D^-)}{I(2D^-)+I(2D^+)}$$

where $I(2D^+)$ and $I(2D^-)$ denote the maximum intensities of the peaks at 2706 cm$^{-1}$ and 2685 cm$^{-1}$ respectively. $\Theta$ is a useful indicator of the expected coverage obtained in our experiments where oxidation proceeds at 40°C (Fig. 2b) Moreover, $\Theta$ is a better indicator of the coverage than the oxidation time, as the obtained coverage saturates for times over 120 minutes in this case when the sample is fully oxidized and $\Theta$ approaches 1. We note that in cases where the oxidation of copper is performed at room temperature the graphene coverage does not approach 100% until $\Theta$ reaches 1, i.e. until the complete disappearance of the 2D$^+$ peak (Supplementary Fig. S2, Discussion).

Optical images of graphene on Cu foil before and after immersion in water illustrate how the Cu oxidation is initiated along lines on the Cu surface (Fig. 2c-d). A more homogeneous oxidation of the surface first occurs after prolonged oxidation time (Fig. 2e).

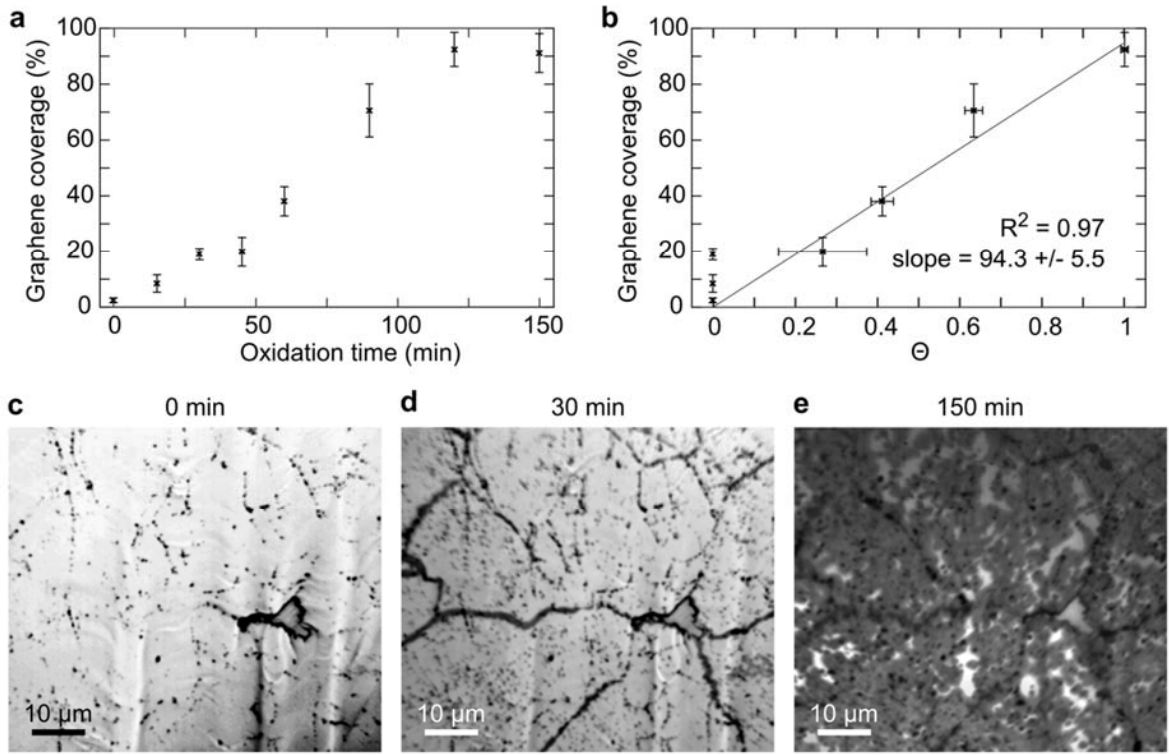

**Fig. 2.** (a) Graphene coverage on SiO$_2$ after transfer as function of the oxidation time in DI water at 40°C. (b) Graphene coverage on SiO$_2$ after transfer as a function of Θ. (c) Optical image of Cu/graphene before immersion in water. (d,e) Optical images of Cu/graphene after immersion in water for 30 and 150 minutes, respectively.

Graphene grown on thin-film catalyst layers of up to 300 mm diameter display a homogeneous monolayer characteristic which is maintained after transfer by this technique, with over 99.7% of the mapped areas showing monolayer graphene contrast (Fig. 3a). The remaining 0.3% consists in the majority of holes, resulting either from untransferred graphene or holes in the as-grown graphene layer. The quality of the graphene as determined from distributions of the Raman I(D)/I(G) peak intensity ratios and the 2D peak full width at half maximum (FWHM), Γ$_{2D}$, at 100 individual random points on the sample surface reveal very low defect density with the mean I(D)/I(G) < 0.058 ± 0.025 and Γ$_{2D}$ of 36.5 ± 1.7 cm$^{-1}$, which is consistent with previously reported values for graphene transfers (Fig. 3b-d).[19,31,32]

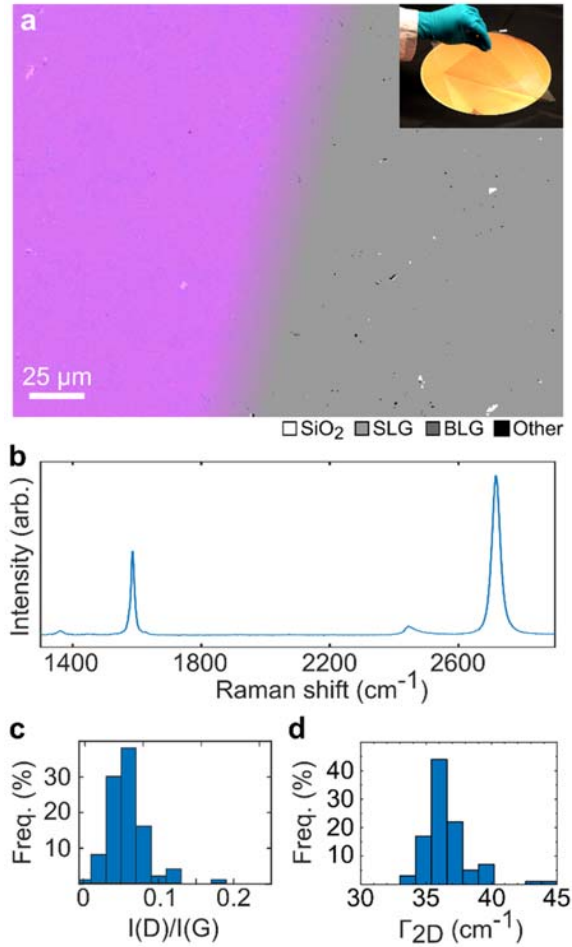

**Fig. 3.** Graphene transferred from thin film Cu on a 12 inch Si wafer. (a) White light optical microscopy (left) and coverage image (right) of graphene transferred to $SiO_2$. Inset shows a photograph of graphene being transferred from a 12-inch wafer using PVA. (b) A representative Raman spectrum from the sample shown in (a). (c,d) Histograms of I(D)/I(G) peak ratios and $\Gamma_{2D}$.

Cross-sectional samples of the graphene on catalyst structure after water oxidation were prepared by FIB and studied via TEM. Water oxidized samples show a uniform 4.3 ± 0.8Å layered structure (Fig. 4a,b) which corresponds well to the cubic $Cu_2O$ lattice parameter of 4.26 Å,[33] but with a relatively large variation across the interface reflected in the spread of this measurement. In addition, we observe voids up to 100 nm in diameter in the catalyst layer (Fig. 4c) in samples oxidized for > 10 hours.

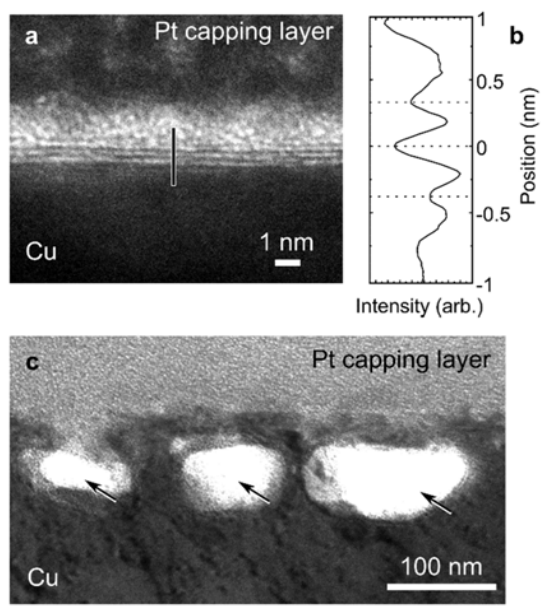

**Fig. 4.** (a) TEM image of Cu/graphene interface after water oxidation showing a layered structure. (b) Line profile for the line across the interface in (a). (c) TEM image of Cu/graphene interface after water oxidation with arrows highlighting voids in the catalyst layer.

We note that the choice of polymer plays a role in the transferred material coverage, with PVA transfers providing the highest coverage here. We also tested poly(methyl methacrylate), polypropylene carbonate, polyvinyl butyral, cellulose acetate butyrate, and polyvinylpyrrolidone as support layers for the transfer with otherwise identical transfer procedures, but none resulted in coverages as high as those obtained for PVA-based transfers (Supplementary Fig. S3).

Additionally, we successfully transferred large areas of single layer CVD-grown hBN from commercially available Cu foil. XPS was used to confirm the expected stoichiometry of hBN on Cu before oxidation (Supplementary Fig. S4). It has previously been shown that ambient oxygen can intercalate and oxidize the interface between hBN and Cu.[34] A sample was oxidized in water at 40°C for 24 hours and the hBN was subsequently transferred using PVA. Optical and coverage images of hBN transferred to $SiO_2$ are shown in Fig. 5a. The hBN to $SiO_2$ coverage of the sample is 97.5%. The hBN Raman peak at 1369 $cm^{-1}$ is in good agreement with the expected value for single layer hBN (Fig. 5b).[25,35] Transfers of hBN were carried out after different immersion times in water at 40°C. There was a clear difference in the hBN coverage on $SiO_2$ after transfer for 90 minutes and 120 minutes of immersion (Supplementary Fig. S5).

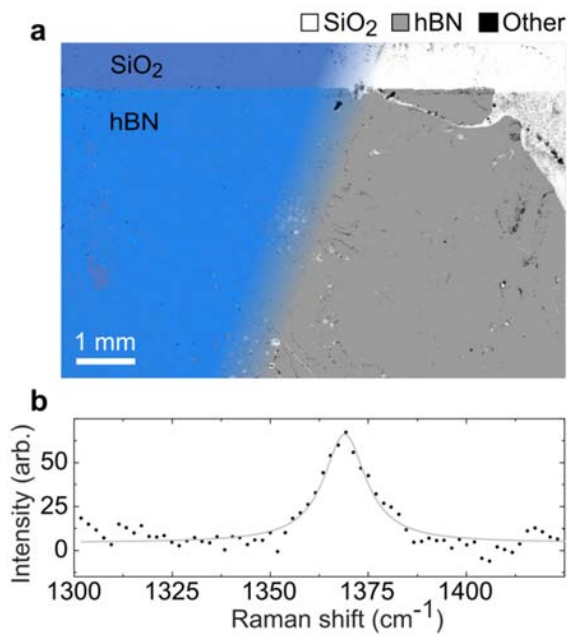

**Fig. 5.** (a) White light optical microscopy (left) and coverage map (right) of hBN on $SiO_2$. (b) Raman spectrum of single layer hBN on $SiO_2$. Points indicate the raw data, while a Lorentzian fit is shown as a solid line.

## 4. Discussion

Cu$_2$O and CuO are formed naturally when Cu is exposed to dissolved oxygen according to the reactions[17]

$$2Cu + O_2 + 2H_2O \rightarrow 2Cu(OH)_2 \text{ and } 3Cu(OH)_2 \rightarrow CuO + Cu_2O + 3H_2O + \tfrac{1}{2}O_2.$$

These reactions occur at imperfections in CVD grown graphene (and hBN) such as wrinkles, grain boundaries, atomic defects, and directly exposed Cu areas due to incomplete growth, and subsequently spreads underneath the graphene covered areas.[36–39]

Raman spectrometry samples a small region of the surface of the sample defined by the laser spot size (typically less than 1 µm FWHM if a 50x or above objective is used), which can contain both oxidized and unoxidized regions. Spatially inhomogeneous decoupling of the graphene from the substrate is the most probable source of the simultaneous presence of two 2D peak components. As a result, monitoring for the complete disappearance of the 2D$^+$ peak and its substitution with the 2D$^-$ peak, i.e. when $\Theta = 1$, enables the point of complete decoupling of the graphene from the surface to be precisely determined. We note that in some cases, samples can acquire an oxide layer under ambient conditions: that is, through contact with atmosphere. In these cases a short duration water oxidation does not lead to the degree of coverage that would be expected from the value of $\Theta$ (Supplementary Fig. S2). It is therefore likely that the oxidation of the copper catalyst layer is a necessary but not sufficient criterion for the complete transfer of graphene.

TEM evidence of catalyst surface pitting for water oxidized samples provides some insight here. In order to dissolve the catalyst layer and produce voids over the surface, there must be good exchange of electrolyte between the water bath and the pitting region. This implies that there is a layer of intercalated water between the catalyst and graphene which mediates this pitting and assists in decoupling the graphene from the substrate. In this instance, the graphene acts as an electrode and provides one half of a galvanic cell which enables this pitting to take place, albeit more slowly in deionized water.

Clearly, such galvanic reactions are not required for the successful complete transfer of CVD grown 2D materials, as demonstrated by our transfer of CVD grown and non-conductive hBN. Here, only the oxidation of the catalyst layer and water intercalation can play a role in decoupling the 2D material from the catalyst substrate.

The Raman monitoring technique we present is applicable to any transfer where graphene is delaminated from the catalyst surface by overcoming adhesion forces directly, as opposed to by dissolution of the catalyst chemically or electrochemically. The monitoring of transfer-readiness of multilayer graphene from, e.g. nickel catalysts and other 2D materials is complicated by the Raman response of layers within the bulk which are not in direct contact with the substrate and are not modified by decoupling, and in this study by the ability to achieve partial transfers of the thickness of such multilayers by mechanical exfoliation when peeling, leaving some layers behind on the growth substrate.

$\Theta$ is particularly useful for monitoring the progress of the decoupling for graphene, however hBN does not display any Raman spectral peaks with a similar characteristic splitting behavior, so the readiness for transfer must be inferred using graphene-based calibration of time and temperature of intercalation. Other 2D materials may display unambiguous spectral characteristics of decoupling from growth catalysts similar to graphene. Monitoring of the copper oxide Raman spectral peaks themselves unfortunately does not provide an unambiguous determinant of the degree of decoupling achieved, since a complementary spectral component that decreases with increasing decoupling is also ideally required.

## 5. Conclusion

In conclusion, we have systematically studied the relationship between the evolution of the Raman spectra of graphene on Cu during water oxidation and the graphene coverage after transfer. Our results show that it is possible to determine at which point the graphene is sufficiently decoupled from the Cu substrate to be transferred, which in our case led to coverage of up to 99.7%. Changes in the Raman 2D peak characteristics ($\Theta$) give a reliable indication of the decoupling time, i.e. the time required for the oxidation process to fully decouple the graphene layer. We expect that Raman spectroscopy could consequently be used both for detecting the decoupling time before transfer and for post-transfer characterization.

**Acknowledgements**
P.R.W and A.C.S. acknowledge financial support from Innovation Fund Denmark Da-Gate 0603-005668B. P.R.W., B.L., A.J., L.P. and T.J.B. acknowledge financial support from EU FP7-604000 'GLADIATOR'. P.B. and B.S.J. thank the Danish National Research Foundation Centre for Nanostructured graphene, DNRF103, and EU Horizon 2020 'Graphene Flagship' 696656. R.W. acknowledges EPSRC Doctoral Training Award (EP/M506485/1). L.C. acknowledges funding from the People Programme (Marie Curie Actions) of the European Union Horizon2020 (H2020-MSCA-IF-2014) under grant agreement 658327, 2D Hetero-architecture.

**Appendix A. Supplementary data**

Supplementary data associated with this article can be found, in the online version, at …

Fig. S1 shows Raman spectra acquired during the oxidation process from a sample being oxidized in water at 40°C. Six Raman peaks (215 cm$^{-1}$, 298 cm$^{-1}$, 416 cm$^{-1}$, 499 cm$^{-1}$, 648 cm$^{-1}$, and 795 cm$^{-1}$) appear over time after immersion in water. There is a no apparent D peak in the sample which indicates that the process is non-destructive towards the graphene film.

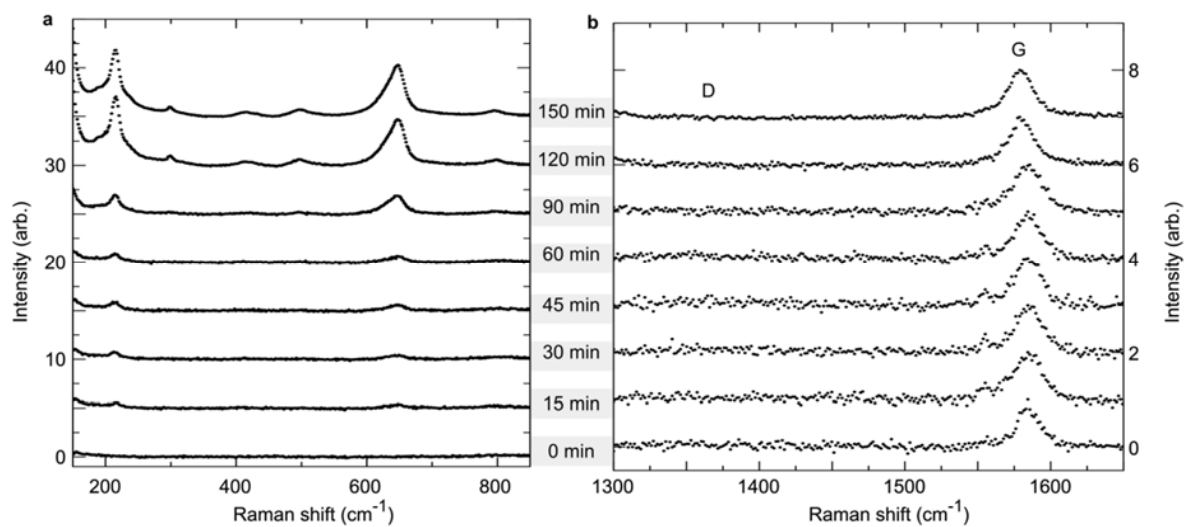

**Fig. S1.** a) Raman Cu oxide peaks and b) D and G peaks acquired from Cu/graphene samples as function of oxidation time in 40°C water. The intensities are normalized to the G peak.

The shift in the position of the 2D peak of graphene to lower wave numbers with increasing oxidation time observed in Fig. 1h also takes place when the Cu oxidation is carried out in DI water at different temperatures. The increase in graphene coverage after transfer with oxidation time for different temperatures is shown in Fig. S2a. It is seen that the process is slower in 25°C water compared to 50°C and 75°C water. Fig. S2b shows the evolution of Θ for Cu/graphene oxidized in DI water at 25°C, 50°C, and 75°C. It is again seen that the graphene coverage reaches its maximum when the $2D^+$ peak vanishes just as the case for 40°C water.

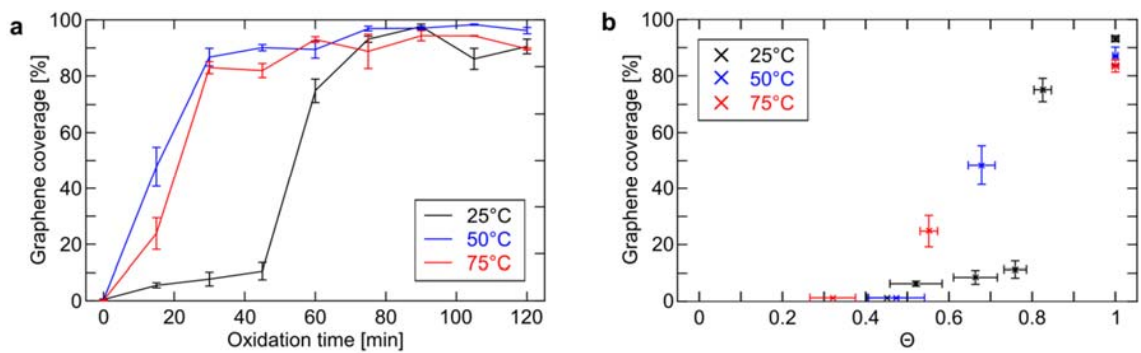

**Fig. S2.** (a) Graphene coverage on $SiO_2$ after transfer as function of the oxidation time in DI water at different temperatures. (b) Graphene coverage on $SiO_2$ after transfer as a function of Θ.

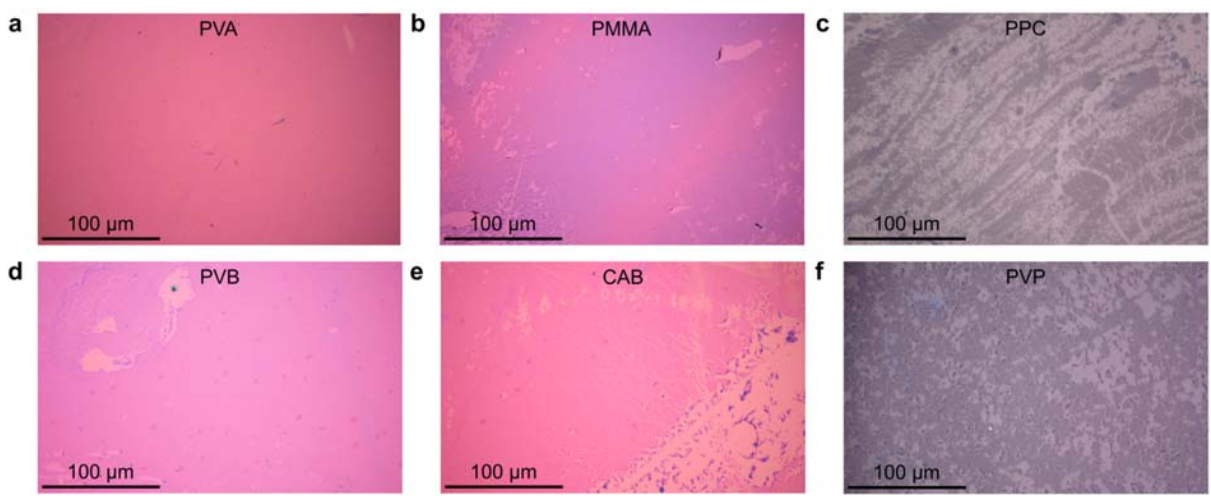

**Fig. S3.** Water-based transfers with different polymers for mechanical peeling of graphene from oxidized Cu. (a) PVA, (b) PMMA, (c) PPC, (d) PVB, (e) CAB, and (f) PVP. The results from PMMA peeling are similar to the soak-and-peel delamination shown by Gupta et al.[16]

Not only is the pre-oxidation of the Cu substrate important, the choice of polymer for the mechanical peeling step also plays an important role. For comparison, other polymers were tested as alternatives to PVA to investigate the influence of the polymer. Experiments were conducted with poly(methyl methacrylate) (PMMA, 8wt% in anisole), polypropylene carbonate (PPC, 8wt% in anisole), polyvinyl butyral (PVB, 8wt% in ethanol), cellulose acetate butyrate (CAB, 20wt% in anisole), and polyvinylpyrrolidone (PVP, 5wt% in DI water) following the same steps as shown in Fig. 1 using a different polymer in the drop coating step and removing the polymer with acetone instead of water (except for PVP which is also removed by water) as the final step.

Although PMMA functions well for wet transfer methods, it is not ideal for transferring graphene continuously by mechanical peeling as shown in Fig. S3 which compares peeling transfers with the six different polymers. It is seen that PVA is superior to PMMA, PPC, PVB, CAB, and PVP with respect to the continuity of the graphene after transfer to $SiO_2$.

Fig. S4a shows a Raman spectrum of Cu with hBN after 1 day in water at 40°C. It is seen that Cu oxide Raman peaks also occur under CVD hBN after leaving the samples in water as in the case of graphene. The hBN Raman signal is very weak on Cu, so only the Cu oxide peaks are visible with the used collection parameters. XPS of hBN on Cu before water oxidation is shown in Figure S4b confirming the presence of hBN on the sample.

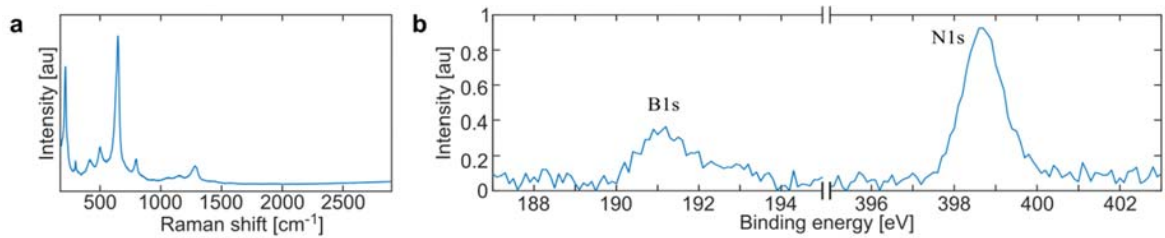

**Fig. S4.** (a) Raman spectra of Cu with hBN after 1 day in DI water at 40°C. (b) XPS peaks for B and N before Cu oxidation.

Fig. S5 shows optical and coverage images of hBN transferred to SiO$_2$ after 90 minutes and 120 minutes immersion in DI water at 40°C. The hBN coverage after 90 minutes immersion is 42.7%, which is much lower than the coverage of 98.5% obtained after 120 minutes of immersion.

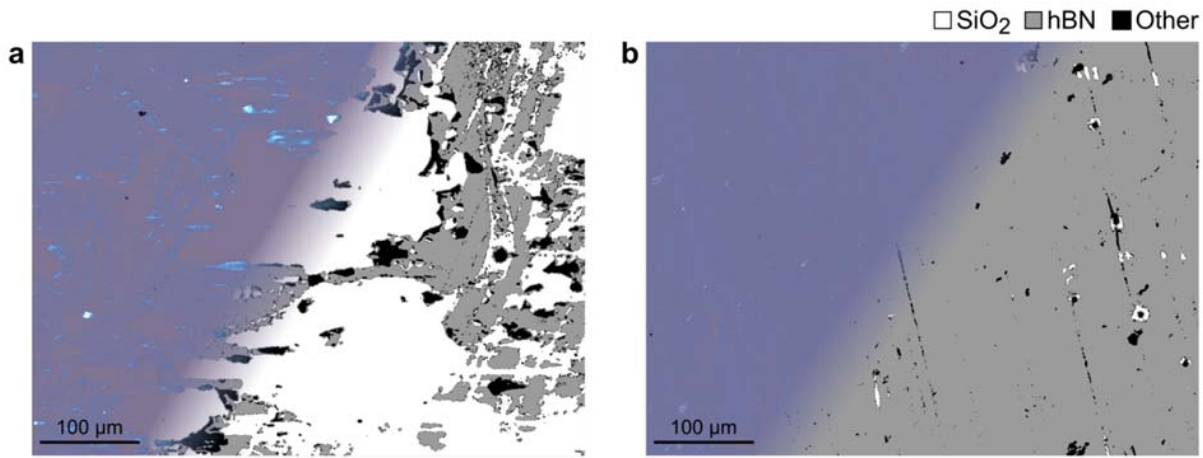

**Fig. S5.** White light optical microscopy (left) and coverage map (right) of hBN transferred to SiO$_2$ after (a) 90 minutes and (b) 120 minutes immersion in DI water at 40°C.